\documentclass[11pt]{article}
\usepackage{amssymb,amsfonts,amsmath,longtable,lscape,graphicx}

\newcommand{\be}{\begin{eqnarray*}}
\newcommand{\ee}{\end{eqnarray*}}
\newcommand{\ben}{\begin{eqnarray}}
\newcommand{\een}{\end{eqnarray}}
\newtheorem{prop}{Proposition}
\newcommand{\bprop}{\begin{prop}}
\newcommand{\eprop}{\end{prop}}
\newtheorem{proof}{Proof}
\newcommand{\bproof}{\begin{proof}}
\newcommand{\eproof}{\end{proof}}
\newtheorem{thm}{Theorem}
\newcommand{\bthm}{\begin{thm}}
\newcommand{\ethm}{\end{thm}}

\def\fr{\mathfrak{f}}
\def\lae{E}
\def\lak{{\mathfrak{k}}}
\def\lap{{\mathfrak{n}}}
\def\csa{{\mathfrak{h}}}
\def\half{\frac{1}{2}}
\def\reals{\mathbb{R}}
\def\ints{\mathbb{Z}}

\def\lag{{\mathfrak{g}}}
\def\mult{\mbox{mult}}
\def\weyl{{\mathcal{W}}}
\def\funddom{{\mathcal{C}}}
\def\suppfun{{\mathcal{K}}}

\begin{document}

\begin{titlepage}
\begin{flushright}
hep-th/0304246\\
DAMTP 2003-35\\[2cm]
\end{flushright}
\begin{center}
{\bf \Large $\lae_{11}$ as $\lae_{10}$ representation at low levels}\\[2cm]
{\large Axel Kleinschmidt\\}
Department of Applied Mathematics and Theoretical Physics\\
University of Cambridge, Wilberforce Road, Cambridge CB3 0WA, UK\\
Email: {\tt A.Kleinschmidt@damtp.cam.ac.uk}\\[1cm]
\mbox{}\\[2cm]

\end{center}

\renewcommand{\abstract}{\begin{center}\bf
Abstract\\[.3cm]\end{center}}

\begin{abstract}
The Lorentzian Kac-Moody algebra $\lae_{11}$, obtained by doubly
overextending the compact $\lae_{8}$, is decomposed into representations
of its canonical hyperbolic $\lae_{10}$ subalgebra.  Whereas the
appearing representations at levels $0$ and $1$ are known on general
grounds, higher level representations can currently only be obtained
by recursive methods. We present the results of such an analysis up to
height $120$ in $\lae_{11}$ which comprises representations on the first
five levels. The algorithms used are a combination of Weyl orbit
methods and standard methods based on the Peterson and Freudenthal
formulae. In the appendices we give all multiplicities of $\lae_{10}$ occuring
up to height $340$ and for $\lae_{11}$ up to height $240$. 
\end{abstract}

\end{titlepage}

\begin{section}{Introduction}

Starting from the largest exceptional classical Lie algebra
$\lae_8$ one obtains the affine $\lae_9$, the hyperbolic $\lae_{10}$
and the Lorentzian $\lae_{11}$ Kac-Moody (KM) algebras\footnote{We use
the notation $\lae_k$ rather than $\mathfrak{e}_k$ in concordance with
most of the literature on the subject.} by extending the
Dynkin diagram \cite{GoOl85,Ka90,KaMoWa88}. The finite-dimensional 
$\lae_8$
is  rather well understood and plays an important
r\^ole in the construction of heterotic string theories. Affine
$\lae_9$ is also under control and appears in supergravity or
Wess-Zumino-Witten 
models, but $\lae_{10}$
and $\lae_{11}$ are far less well understood. $\lae_{8}, \lae_9$ and
$\lae_{10}$  have been
suggested as symmetries of supergravity when dimensionally reduced to
$3,2$ and $1$ dimension \cite{Ju82,MaSc83}. 
$\lae_{10}$
has also appeared recently in cosmological billiards for describing
gravitational systems coupled to $p$-form and scalar matter 
near a spacelike singularity
\cite{DaHeNi02a,DaHeNi02b}. 
On the other hand, $\lae_{11}$
has been proposed as a (hidden) 
symmetry of the full eleven-dimensional theory and also of M theory
\cite{We01,We02} and belongs to the class of very extended algebras
which have been considered in \cite{GaOlWe02,EnHoTaWe03}. 

There have been a number of attempts at understanding hyperbolic KM
algebras (see for instance \cite{FeFr83,FeFrRi93,GeNi95}) but to date
not a single example has been understood in full detail. In this paper
we also only present partial results, partly obtained with the help of
a computer. These results are mainly for the physically
interesting cases of $\lae_{10}$ and $\lae_{11}$ but also offer some
general observations for arbitrary (indefinite) Kac-Moody algebras
and results for the rank $2$ case. 

As $\lae_{10}$ is naturally a subalgebra of $\lae_{11}$ we consider
the problem of decomposing $\lae_{11}$ with respect to this
subalgebra. This is a somewhat different approach from
\cite{NiFi03} where both algebras were decomposed at low levels
with respect to their $A_9$ and $A_{10}$ subalgebras and from
\cite{KaMoWa88} where a decomposition of $\lae_{10}$ with
respect to its affine $\lae_9$ algebra was considered. Our approach is
to consider hyperbolic sections of the $\lae_{11}$ root lattice rather
than the elliptic or affine slicings considered so far. This will give
us information about the algebra deep inside the light-cone, at least
if we know which representations occur. It can be shown that the
representation on level
$1$ is the highest weight representation of $\lae_{10}$ 
with highest weight $\lambda_{-1}$ by generalising results from
\cite{FeFr83,FeFrRi93}.  
For other levels, we find representations of $\lae_{10}$ with 
various outer multiplicities. Our methods enable us to calculate
the complete structure of $\lae_{11}$ as seen by
$\lae_{10}$ up to height 120 which corresponds to some of the
low-lying representations up to level $5$. Compared to the elliptic
approach this calculation does not make direct contact with fields
from supergravity but aims primarily at revealing some of the structure
of $\lae_{10}$ and $\lae_{11}$. Even though neither $\lae_{10}$ nor
$\lae_{11}$ are understood fully, we might still learn something
about them by studying their interrelation or combining different 
decompositions of the same structure. The calculation done in this
paper is the first explicit calculation of large parts of $\lae_{10}$
representations and can be used for instance to obtain lower bounds
for $\lae_{11}$ multiplicities.
We also present raw
data for both algebras separately in two appendices. There it is 
noted that the multiplicity of a root compared to its light-cone
degrees of freedom behaves rather differently for $\lae_{10}$ and
$\lae_{11}$, supporting an observation in \cite{GeNi95,NiFi03}.
 It would be good to have a conceptual understanding
of this behaviour.

Whereas the hyperbolic slicing of $\lae_{11}$ is rather unproblematic,
it is not clear how to obtain the relevant representations of
$\lae_{10}$ with respect to one of its hyperbolic (or even Borcherds)
subalgebras. The reason is that the representations are not of highest
weight type and it would be interesting to understand their structure
better. 

The structure of this note is as follows. Section 2 contains some of
the background we are using on Kac-Moody algebras and on their
decompositions, illustrated by a rank 2 example. 
In section 3 we outline the method used to obtain the
decomposition of $\lae_{11}$ with respect to $\lae_{10}$ and list the
appearing representations. In appendices we give complete tables for
the multiplicities of the imaginary 
roots in the fundamental domains of $\lae_{10}$
and $\lae_{11}$, most of which have not appeared in the literature
before.

\end{section}

\begin{section}{Notation}

Our notation is mostly borrowed from Kac' book \cite{Ka90}. Given a
symmetric, non-degenerate\footnote{We restrict to the case of
non-degenerate Cartan matrices all the way through the article as
this is the only case we will be concerned with. We also assume the
matrix to be indecomposable. The symmetry of $A$ implies
$a_{ii}=\alpha_i^2=2$ for all $i$ 
and the invariant bilinear form is just given by $A$.}
Cartan matrix $A=(a_{ij})_{i,j\in I}$, the collection of simple
roots $\alpha_i$ of a Kac-Moody algebra $\mathfrak{g}$ is  denoted $\Pi$,
where $i$ belongs to some (finite) index set $I$, and the simple roots
generate the root lattice $Q$ freely. The number of simple roots is
called the rank of the KM algebra. The simple co-roots
$\Pi^\vee=\left\{\alpha^\vee_i\right\}$ 
are a basis for the Cartan subalgebra $\mathfrak{h}$ and satisfy
$\alpha_i(\alpha^\vee_j)=a_{ji}$. The
Kac-Moody algebra $\mathfrak{g}$ 
is then defined by the following relations on the (Chevalley) generators
$\alpha_i^\vee,e_i,f_i$ as follows
\be
\left[e_i,f_j\right]&=&\delta_{ij}\alpha^\vee_i,\\
\left[\alpha^\vee_i ,e_j\right]&=&\alpha_j(\alpha^\vee_i)e_j,\\
\left[\alpha_i^\vee ,f_j\right]&=&-\alpha_j(\alpha^\vee_i)f_j,\\
(\mbox{ad}\,e_i)^{(1-a_{ij})}e_j&=&0,\\
(\mbox{ad}\,f_i)^{(1-a_{ij})}f_j&=&0.
\ee
Then it is one of the basic properties of $\lag$ that it has a
triangular decomposition of the form
\ben
\label{triang}
\lag=\lap_+\oplus\csa\oplus\lap_-
\een
where the Cartan subalgebra $\csa$ is
the space spanned by the simple co-roots because we are dealing with
non-degenerate Cartan matrices. The dimension of $\csa$ is equal to
the rank of $\lag$.  
The positive part $\lap_+$ and the negative part $\lap_-$ are
exchanged by the Chevalley involution. So it is sufficient to study
either $\lap_+$ or $\lap_-$ when trying to understand the structure of
$\lag$. Central objects in the study of $\lag$ are those non-zero elements
$\alpha\in\csa^*$ for which there exist elements $x\in\lag$ such that
$\left[h,x\right]=\alpha(h)x$ for all $h\in\csa$. Such elements
$\alpha$ are called roots and the corresponding space of all
$x\in\lag$ satisfying this condition is called the root space
$\lag_\alpha$. 
The number of
independent elements in the root space is called the multiplicity of
the root and denoted by $\mult(\alpha)=\mbox{dim}\lag_\alpha$. The set
of all roots is called $\Delta\subset Q$. From the 
triangular decomposition (\ref{triang}) we deduce
that every root is either positive or negative and so
$\Delta=\Delta_+\uplus\Delta_-$. 

The reflections in the simple co-roots $\alpha^\vee_i$ 
generate the Weyl group $\weyl$ 
which is a Coxeter
group with relations given by the Cartan matrix (or the corresponding
Dynkin diagram). Define the fundamental domain
$\funddom\subset\csa^*$ 
as the intersection of all half spaces which take non-positive values
on a simple co-root, viz.
\be
\funddom=\left\{\alpha\subset\csa^*: \alpha(\alpha^\vee_i)\le
0,\,\mbox{for all simple co-roots}\,\alpha^\vee_i\right\}.
\ee
(In the case of finite-dimensional $\lag$, $\funddom$ is a fundamental
domain for the action of $\weyl$ on $\csa^*$. Our conventions are
adapted to the hyperbolic case defined below.) We define the Weyl
vector $\rho\in\csa^*$ by $\rho(\alpha^\vee_i)=-1$ for all simple
co-roots and so $\rho$ lies in the fundamental domain. Then the set of
(positive roots) and their multiplicities can be deduced in principle
from the orbit of $\rho$ under $\weyl$ by the denominator formula
\cite{Ka90} 
\ben
\label{kmden}
\prod_{\alpha\in\Delta_+}(1-e^{-\alpha})^{\mbox{mult}(\alpha)}=\sum_{w\in W}
\epsilon(w)e^{w(\rho)-\rho}.
\een
The set of roots $\Delta$ has two distinct parts, namely it
splits into real and imaginary roots
\be
\Delta=\Delta^{real}\uplus\Delta^{imag},
\ee
where $\Delta^{real}=\weyl(\Pi)$ and
$\Delta^{imag}=\weyl(\suppfun)\uplus\weyl(-\suppfun)$. Here
$\suppfun=\left\{\alpha\in\funddom:\right. $ supp$(\alpha)$ is
  connected $\left.\right\}$. We have $\alpha^2=(\alpha|\alpha)=2$ for real
roots and $\alpha^2\le 0$ for imaginary roots. The set of roots
(together with their multiplicities) is Weyl invariant. The
denominator formula (\ref{kmden}) can be used to determine the roots
and their multiplicities by just expanding both sides of (\ref{kmden})
or using
recursion identities derived from (\ref{kmden}) like the Peterson formula.
We define the fundamental weights $\lambda_i\in\csa^*$ ($i\in I$) as minus the
dual basis to the simple roots with respect to $A$, 
i.e. they satisfy $(\alpha_i|\lambda_j)=-\delta_{ij}$. We sometimes
write a root $\alpha=\sum n_i \alpha_i=\sum l_i \lambda_i$ as
$(n_1,...,n_{\mbox{\footnotesize{rk}}(\lag)})$ in the simple root
basis or $[l_1,...,l_{\mbox{\footnotesize{rk}}(\lag)}]$ in the
fundamental weight basis. The $l_i$ are called Dynkin labels of
$\alpha$. From the definitions of $\funddom$ and the fundamental
weights we see that their positive linear combinations generate
$\funddom$. 

If $A$ is a positive definite matrix the corresponding KM algebra will
be one of the classical series and in particular
finite-dimensional. If $A$ is positive semi-definite and has exactly
one zero eigenvector the corresponding algebra is affine. 
We call a KM algebra
hyperbolic if its Cartan matrix is indefinite and 
if any matrix obtained from its Cartan matrix by deleting
one row and column is the Cartan matrix of a finite-dimensional or
affine KM algebra. It is known that the rank 
of the algebra cannot exceed $10$ for a hyperbolic KM
algebra \cite{Ka90}. Among the hyperbolic KM algebras of maximal
rank $\lae_{10}$ is distinguished by being the only overextended
one and having as its root lattice the unique even and self-dual
ten-dimensional Lorentzian lattice ${\rm II}_{9,1}$.
The signature of the Cartan matrix of 
any hyperbolic algebra of rank
$n$ is $n-2$. For hyperbolic Kac-Moody algebras we also have
$\Delta=\left\{\alpha\in
Q:\alpha^2\le 2\right\}\backslash\left\{0\right\}$ (see
\cite{Mo79}) and since $A^{-1}$
has only negative entries all $\lambda_i$ are non-negative linear
combinations of the simple roots and so $\funddom\subset
Q_+\otimes\reals$. It has also been established that every hyperbolic
KM algebra has a principal $SO(1,2)$ subalgebra \cite{NiOl01}.
 We call a KM algebra Lorentzian if the Cartan matrix has
signature $n-2$. So any hyperbolic KM algebra is Lorentzian but the
converse is not necessarily true.
There is a well-known procedure for obtaining non-twisted 
affine or hyperbolic
algebras by extending the Dynkin diagrams of the finite cases
\cite{GoOl85}. Extending the diagram again gives a Lorentzian KM
algebra \cite{Ru92,GaOlWe02}.
Highest weight representations with highest weight
$\Lambda\in\funddom$ 
are
denoted by $L(\Lambda)$ and are subject to the Weyl-Kac character
formula.\footnote{Strictly speaking, we are considering lowest weight
representations due to our conventions.}

\begin{subsection}{Subalgebra decompositions}

Suppose we have a subalgebra $\lak\subset\lag$ then $\lag$ is a
representation of $\lak$ under the adjoint action and we can try to
decompose $\lag$ as a sum of representations of $\lak$. The simplest
cases occur when we consider subalgebras that are obtained by deleting
a number of roots from the Dynkin diagram of $\lag$ but these are by
far not all. 
The modules with respect to such an algebra at a
given level\footnote{The level is an element of the quotient of the
root lattices.} are integrable and belong to the category
$\mathcal{O}$ \cite{Ka90}. Thus, they can be
decomposed into a sum of irreducible highest weight representations.
The simplest example is
removing all but one point and we are left with a decomposition with
respect to
one of the standard $\mathfrak{sl}_2$ subalgebras. 

Hyperbolic KM algebras obtained by overextension of an affine
algebra can be decomposed with respect to the affine subalgebra given
by deleting the overextended root and for low levels this yields
explicit formulae for the multiplicities of the roots at this level
\cite{FeFr83,KaMoWa88}. 

In general, removal of a single node from the Dynkin diagram is
governed by the following proposition which is an application of
results from \cite{FeFr83,KaMoWa88,FeFrRi93}.

\begin{prop}
\label{ff}
Let $\lag$ be a Kac-Moody algebra of rank $n$ 
given by Cartan matrix $A$ with 
associated Dynkin diagram. If we remove a single node,
$\alpha_1$ say,
from the Dynkin diagram and define the level $l$ of a root $\alpha$ of
$\lag$ to be the number of $\alpha_1$ summands of $\alpha$. Denote the
index set for the simple roots connected to $\alpha_1$ by $J=\{j\in\{2,..,n\}:
a_{1j}\ne 0\}$.  Then we have the following decomposition of $\lag$ with 
respect to the Kac-Moody subalgebra
associated with the Cartan matrix obtained by deleting the first row
and column
\be
\lag=\bigoplus_{l\in\ints}\lag_l
\ee
where for $l\ne 0$
\be
\lag_l\cong\bigoplus_{i=1}^{N_l}L(\Lambda_i^{(l)})^{\oplus\mu(\Lambda_i^{(l)})}.
\ee
The $N_l$ can be infinite. The $\lag_l$ are given in more detail by:

${\bf l=0}$: The direct sum of the adjoint of the smaller algebra
with a trivial representation. If the smaller algebra is affine then
the trivial representation is absent.

${\bf l=1}$: The highest weight representation of the smaller algebra
with highest weight given by the sum of the fundamental weights
corresponding to the simple roots connected to $\alpha_1$,
i.e. $g_1\cong L(\sum_{j\in J} \lambda_j)$.

${\bf l>1}$: Now consider the free algebra on the roots at $l=1$ and
call it $\fr$ which is also graded by $l$. 
Within it there are the elements $ad^{1+a_{1j}}e_1 e_j$
which are non-zero in the free algebra but vanish in the Kac-Moody
algebra. At the same time they are highest weight vectors for the smaller
algebra and generate an irreducible representation $U_j$ at
$l=1+a_{1j}$. Consider the ideal $\mathfrak{i}_j$ generated within the
free algebra by $U_j$ and define $\mathfrak{i}=\oplus_j
\mathfrak{i}_j$. Then the quotient $\fr/\mathfrak{i}$ describes the
positive part of the $\lag$ for $l>1$.

${\bf l<0}$: These cases are obtained by the Chevalley involtuion from
the cases above.
\end{prop}

The proof of this statement is a simple adaptation of arguments in
\cite{FeFr83,KaMoWa88,FeFrRi93}. The proposition implies that
we might as well recursively use the above decomposition to obtain
information about the Kac-Moody algebra of interest. The problem of
understanding the KM algebra $\lag$ is then reduced to describing the
ideal because the structure of $\fr$ is well-known (for
finite-dimensional first level at least): The free 
algebra on $m$ generators $\beta_k$ ($k=1,..,m$) is governed by
the equation

\ben
\label{denfree}
\prod_{\beta>0}(1-e^\beta)^{\mbox{mult}(\beta)}=1-\sum_{k=1}^m
e^{\beta_k}
\een
from which one can deduce the Witt formula for the dimension of the
graded pieces by M\"obius inversion after specialisation. For more on
free algebras see \cite{Vi78,KaKi96}. What we will also need here
is the multiplicity of a root $\beta=\sum_k n_k\beta_k$ 
in the free algebra obtained by
M\"obius inversion of (\ref{denfree})
\be
\mbox{mult}(\beta)=\sum_{d|(n_1,..,n_m)}\mu(d)\frac{1}{\sum
n_k}
\left(\begin{array}{c}\frac{1}{d}\sum
n_k\\\frac{n_1}{d},..,\frac{n_m}{d}\end{array}\right),
\ee
where 

\be
\left(\begin{array}{c}\sum t_k\\t_1,..,t_m\end{array}\right)
=\frac{(\sum t_k)!}{t_1!..t_m!}
\ee

is a polynomial coefficient. 
We have to note that
several elements in the free algebra will map into the same root space
in the Kac-Moody algebra, where the number of such elements in a given
example can be written easily in terms of a generating function as a
standard combinatorial problem. Despite this complication, it is still possible to determine the
dimension of a root space in the free algebra for a root 
at a given level without prior knowledge of the lower levels,
basically because one knows the multiplicities of the primitive
roots. This is
in contrast to the recursive nature of the Peterson formula for 
multiplicities in the
Kac-Moody algebra where one has a similar expression for the
multiplicities but does not know the answer for the primitive roots
and so is left with calculating recursively until one finally reaches
the root one is interested in. 
(This procedure gets more and more
expensive as one goes up in level in the hyperbolic case as the root
multiplicities and the number of roots are known to increase
exponentially.)

So, if we were able to calculate the dimension of the space associated
with a root $\beta$ in the ideal $\mathfrak{i}$ we could determine the
root multiplicities in the hyperbolic Kac-Moody algebra in a
non-recursive way. It is intriguing to note that from this perspective
there is no difference at all between the finite, affine or
indefinite cases which makes it quite surprising that the finite and
affine (tame) cases are well understood whereas the (wild) indefinite ones
are not. Unfortunately, it is not known how to work out the
ideal in a closed form similar to the free algebra.\\

In order to obtain a feeling for the observations made above we give a few
results for the simplest hyperbolic case which is determined by the
Cartan matrix
\ben
\label{rk2km}
\left(\begin{array}{cc}2&-3\\-3&2\end{array}\right).
\een
This algebra carries some information from the Fibonacci numbers
since the sum over the Weyl group $\mathbb{Z}\ltimes\mathbb{Z}_2$ can
be seen to  be governed by a recursion relation among the odd
Fibonacci numbers \cite{Fe80}.
We denote the roots by pairs $(m,n)$ and present the multiplicities of
some low-lying roots in table \ref{trk2mul}.

\begin{table}
\begin{tabular}{c|rrrrrrrrrrrrrrrrrrr}
$m/n$&0&1&2&3&4&5&6&7&8&9&10&11&12\\
\hline
0&&1&&&&&&&&&&&\\
1&1&1&1&1&&&&&&&&&\\
2&&1&1&2&1&1&&&&&&&\\
3&&1&2&3&4&4&3&2&1&&&&\\
4&&&1&4&6&9&9&9&6&4&1&&\\
5&&&1&4&9&16&23&27&27&23&16&9&4\\
6&&&&3&9&23&39&60&73&80&73&60&39\\
7&&&&2&9&27&60&107&162&211&240&240&211\\
8&&&&1&6&27&73&162&288&449&600&720&758\\
9&&&&&4&23&80&211&449&808&1267&1754&2167\\
10&&&&&1&16&73&240&600&1267&2278&3630&5130\\
11&&&&&&9&60&240&720&1754&3630&6559&10531\\
12&&&&&&4&39&211&758&2167&5130&10531&19022\\
\end{tabular}
\caption{\label{trk2mul} Multiplicities for the rank 2 hyperbolic
KM algebra determined by (\ref{rk2km}).}
\end{table}

Now we decompose this KM algebra with respect to one of its standard
$\mathfrak{sl}_2$-subalgebras and read off the representation content
for the first few rows. The result is given in table \ref{trk2rep} where
we have denoted the $(n+1)$-dimensional representation of
$\mathfrak{sl}_2$ by $P_n$ and the level with respect to that
subalgebra by $l$.

The ideal is generated by the representation $P_{10}$ at
level $4$ in agreement with our general arguments. Trivially, 
at higher levels
the ideal is always contained within the tensor product of $P_{10}$
with the right number of factors of the
first level representation $P_3$. We can say a bit more about the
detailed structure. 

The ideal is precisely given by the tensor product up to $l=8$ when a
defect in this pattern appears around $n=21$ and a reflection
pattern appears in addition to the tensor product structure because
$P_{10}$ at most influences a ``width'' of $21$ in tensor products.
Thus we
can obtain the value for $n=0$ by subtracting the value of $n=22$ from
$n=20$. The reason for this defect is the Serre relation which comes
into action for the first time in a quadruple commutator for our
algebra and so at $l=8$ in the ideal and which reflections appear can
also be understood in this way. 
This reflection behaviour persists up to $l=12$ where it is
being augmented by another reflection property, again deriving from
the Serre relation. Similar patterns appear for the other rank 2 cases
and have been checked to some detail.

Apparently, understanding the precise nature of
these patterns and their occurence is tantamount to understanding the
Serre relations and the 
Jacobi identity in the original KM algebra to all levels. Solving
the ideal problem seems as hard as solving the original problem. \\

Subalgebras obtained by deleting just a number of nodes together with
the edges connected to them are not the
only subalgebras which can appear. For instance, any affine KM
algebra has affine subalgebras from just restricting the horizontal
(level 0) piece to a subalgebra and then affinizing. The corresponding
Dynkin diagram is not a subdiagram of the original one. In general it
will be much harder to describe the representations with respect to
these more involved cases as they do not necessarily belong to the
category of representations which can be decomposed into highest
weight representations at all levels.

As an example we just mention
that the rank 2 algebra discussed above has a hyperbolic rank 2
subalgebra whose first simple root is identical with an original one
and whose second simple root is a Weyl translate of the other original
simple root. The smallest example corresponds to a subalgebra $\lak$
where the simple roots have inner product
$-7$. At level $l=0$ (where we
have a decomposition into the adjoint plus highest weight
representations) 
we can give
a few of the low-lying representations that appear 
\begin{eqnarray}
\lak\oplus L(5\lambda_1+5\lambda_2)\oplus
L(3\lambda_1+12\lambda_2) \oplus L(12\lambda_1+3\lambda_2)\nonumber\\
 \oplus
L(\lambda_1+19\lambda_2)\oplus L(19\lambda_1+\lambda_2)\oplus ..
\end{eqnarray}
It seems tempting to think that if one could determine all the
representations with respect to this somewhat coarser algebra and then
iterate the procedure one might learn something about the original
algebra because the limiting point of this procedure is a free algebra
where one is again able to calculate everything directly.

Similarly, some KM algebras have Borcherds subalgebras which are
generalizations of Kac-Moody algebras \cite{Bo88}. Again
decomposing with respect to them will soon leave the realm of highest
weight representations and is thus subject to the obstructions
mentioned above. Some related ideas have recently appeared in
\cite{FeNi03}.

\begin{landscape}
\begin{center}
\begin{table}
\begin{center}
\begin{tabular}{ccccccccccccccccccc|c|c}
18&17&16&15&14&13&12&11&10&9&8&7&6&5&4&3&2&1&0&$P_n$/$l$&dim\\
\hline
&&&&&&&&&&&&&&&1&&&&1&4\\
&&&&&&&&&&&&&&1&&&&1&2&6\\
&&&&&&&&&&&1&&1&&1&&1&&3&20\\
\hline
&&&&&&&&&&1&&3&&2&&3&&&4&49\\
&&&&&&&1&&3&&5&&7&&7&&4&&5&160\\
&&&&&&3&&6&&14&&16&&21&&13&&7&6&494\\
&&&2&&7&&18&&33&&47&&55&&49&&29&&7&1636\\
\hline
1&&5&&21&&46&&89&&126&&161&&151&&120&&38&8&5410\\

\end{tabular}
\caption{\label{trk2rep}The rank 2 algebra (\ref{rk2km}) seen from one
of its standard $\mathfrak{sl}_2$ subalgebras.}
\end{center}
\end{table}

\begin{table}
\begin{center}
\begin{tabular}{ccccccccccccccccccccccc|c|c}
22&21&20&19&18&17&16&15&14&13&12&11&10&9&8&7&6&5&4&3&2&1&0&$P_n/l$&dim\\
\hline
&&&&&&&&&&&&&&&&&&&&&&&1&0\\
&&&&&&&&&&&&&&&&&&&&&&&2&0\\
&&&&&&&&&&&&&&&&&&&&&&&3&0\\
\hline
&&&&&&&&&&&&1&&&&&&&&&&&4&11\\
&&&&&&&&&1&&1&&1&&1&&&&&&&&5&44\\
&&&&&&1&&2&&3&&4&&3&&2&&1&&&&&6&176\\
&&&1&&3&&6&&10&&12&&12&&10&&6&&3&&1&&7&704\\
\hline
1&&3&&10&&19&&31&&39&&44&&39&&31&&19&&10&&2&8&2750\\
\end{tabular}
\caption{\label{trk2ideal}The ideal $\mathfrak{i}$ described in the
text for the rank 2 algebra (\ref{rk2km}).}
\end{center}
\end{table}
\end{center}
\end{landscape}

\end{subsection}

\end{section}

\begin{section}{Computational results for $\lae_{10}\subset\lae_{11}$}

In this section we apply the considerations from the last section to
the KM algebras $\lae_{10}$ and $\lae_{11}$ which are hyperbolic and
Lorentzian respectively. The Dynkin diagram of $\lae_{11}$ is given in
figure \ref{dynke11} and the diagram for $\lae_{10}$ is obtained by
deleting the node marked by $-2$.

\begin{figure}
\begin{center}
\includegraphics{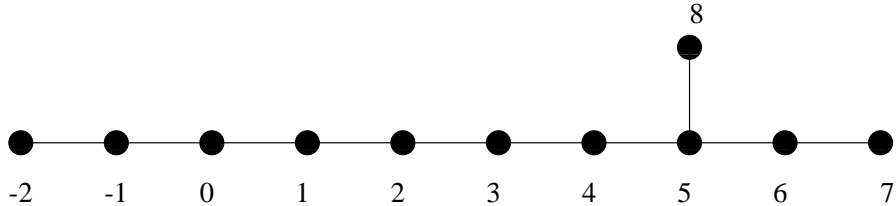}
\caption{\label{dynke11}The Dynkin diagram of $\lae_{11}$.}
\end{center}
\end{figure}

We can apply the denominator directly to obtain multiplicities for
both these algebras and we can consider $\lae_{10}$ as a subalgebra of
$\lae_{11}$ and apply the proposition. In that case we immediately
deduce that the multiplicities of $\lae_{11}$ at level $1$ are given
by the $\lae_{10}$ representation with highest weight
$\Lambda=\lambda_{-1}$, i.e the fundamental weight corresponding to
the overextended node. Then we have
\be
\sum_{\alpha>0}\mbox{mult}(1,\alpha)e^\alpha=\chi_{L(\lambda_{-1})}
\ee
and the multiplicities can be computed from this identity as long as
the $\lae_{10}$ multiplicities are known. This gives information about
$\lae_{11}$ deep inside the forward lightcone. It turns out that doing
the calculation for this representation only is not much faster than
computing the full $\lae_{11}$ up to the same height and so we
consider this problem.

The method we have used for calculating the multiplicities of
$\lae_{10}$ and $\lae_{11}$ relies on the Peterson recursion formula
\ben
\label{petersonrec}
(\beta|\beta+2\rho)c_\beta=\sum_{\alpha>0 :
\beta-\alpha>0}(\alpha|\beta-\alpha) c_{\alpha} c_{\beta-\alpha}
\een
for KM algebras \cite{peterson}. Here the quantities $c_\beta$ are the
expansion coefficients of the negative logarithm of the denominator
(\ref{kmden}). They are related to the multiplicities via
$c_\beta=\sum_k\frac{1}{k}\mult(\beta/k)$. 
By symmetry of the sum one can
restrict to $\alpha$ with height not exceeding half the height of $\beta$.
The algorithmic value of this
relation can be improved by noting that the decomposition
$\beta=\alpha+\alpha'$ is invariant under the stabilizer $\weyl_\beta$
of $\beta$
and so are the inner product and the quantities $c_\beta$. Thus 
we can write the right hand side in (\ref{petersonrec}) as a
sum over orbits under this stabilizer group
\be
(\beta|\beta+2\rho)c_\beta=\sum_{\mbox{\footnotesize orbits}\, o}
(\gamma|\beta-\gamma) c_\gamma
c_{\beta-\gamma} |o|
\ee
where $\gamma$ is any element from the orbit under $\weyl_\beta$ and
$|o|$ the size of the orbit in the positive root domain. If
$\beta=\sum_i l_i\lambda_i$ then the stabilizer is generated by the
reflections in the simple roots $\alpha_i$ for which $l_i=0$. Denote
the set of values of $i$ for which this the case by $J\subset I$.
We can
represent each orbit by a ``lowest weight vector'' $\gamma$ which is
charaterised as having either only non-negative Dynkin labels on the
set $J$ or being a simple root (or a multiple thereof) belonging to
$J$ up to conjugacy. 
The element $\beta-\gamma$ is then a highest element under the
action of $\weyl_\beta$ and if it is not conjugate to a simple root
from $J$ under $\weyl_\beta$ the full orbit will contribute and the
size of the orbit is just the quotient of the order of the
$\weyl_\beta$ by the order of the stabilizer of $\gamma$ within
$\weyl_\beta$.\footnote{This formula does not cause trouble even
though some elements $\beta$ in $\lae_{10}$ and $\lae_{11}$ have stablizers of
infinite order as in that case either the support of $\beta$ is
disconnected or one of $\gamma$ and $\beta-\gamma$ belongs to
$\weyl_\beta\cdot \alpha_j$ for some $j\in J$.}
In the
case that either $\gamma$ or $\beta-\gamma$ is conjugate to a simple
root $\alpha_j$ 
from $J$ the size of the orbit is given by the number of positive
roots of the (finite) subalgebra defined by the connected part of $J$
to which $\alpha_j$ belongs. There is also a possible symmetry factor
of $2$
one has to take into account if the $\gamma$ and $\beta-\gamma$ are
not conjugate under $\weyl_\beta$. This algorithm then is very similar
to the ones described in \cite{MoPa82,BaeGeNi98}. 

In order to illustrate this algorithm we apply it to 
the first non-affine root\footnote{The affine roots all have
multiplicity $8$ because they lie in the $\lae_9$ subalgebra. 
A light-like root thus has fewer polarisations than allowed for
by transversality.} 
of $\lae_{11}$ which is
$\beta=\lambda_7=(1,2,3,4,5,6,7,8,5,2,4)$. Its stabilizer is the Weyl group
of the Lie algebra $\weyl_\beta=\weyl(D_{10})$ of order $2^9 10!$ and the
factor works out to $(\beta|\beta+2\rho)=-96$. We always have the
contributions from $\alpha_7$ and one of the simple roots of $D_{10}$
which are all conjugate under $\weyl_\beta$. Their contributions can
be summarized in the following table:\\

\begin{tabular}{c|c|c|c|c|c|c|c|c|c|c}
$\gamma$&$\gamma^2$&$c_\gamma$&$\beta-\gamma$&$(\beta-\gamma)^2$&
$c_{\beta-\gamma}$&$(\gamma|\beta-\gamma)$&simple&$|o|$&sym&c\\
\hline
$\alpha_7$&2&1&$w(\alpha_7)$&2&1&-3&no&$2^9$&1&-1536\\
$\alpha_6$&2&1&$w(\lambda_{-1})$&0&8&-2&yes&$90$&2&-2880
\end{tabular}\\

Here we have indicated the lowest element in the Weyl orbit of
$\beta-\gamma$. 
The column labelled ``simple'' indicates whether one of the
vectors is conjugate to a simple root of $D_{10}$ under
$\weyl_\beta$. Accordingly, the sizes of the orbits can be computed as
$|\weyl(D_{10})|/|\weyl(A_9)|=2^9 10!/10!=2^9$ in the first case and $90$
is the number of positive roots of $D_{10}$. The column ``sym''
denotes the symmetry factor. 
One can convince oneself that these are all orbits that
contribute. So the multiplicity of $\lambda_7$ is 46 which 
can also be checked independently.\\

We are also interested in highest weight representations of $\lae_{10}$
in order to decompose $\lae_{11}$ along the lines of proposition
\ref{ff}. Denote a highest weight module with dominant highest weight
$\Lambda\in\funddom$ by $L(\Lambda)$ then the Weyl-Kac character formula holds
\cite{Ka90} 
\be
\chi_{L(\Lambda)}=\frac{\sum_{w\in\weyl}\epsilon(w)\exp(w(\rho+\Lambda)-\rho)}
{\prod_{\alpha>0}(1-\exp(-\alpha))^{\mult(\alpha)}}.
\ee
As a consequence, we have the following version of the Freudenthal
formula \cite{Ka90}
\be
(2(\Lambda|\beta)-2\mbox{ht}(\beta)+\beta^2) 
\mult_{L(\Lambda)}(\Lambda+\beta)\\
=2\sum_{\alpha>0}\mult(\alpha)\sum_{j\ge 1}&
(\Lambda+\beta-j\alpha|\alpha)\mult_{L(\Lambda)}(\Lambda+\beta-j\alpha).
\ee
Here the right hand side can again be decomposed into a
sum over Weyl orbits. Using these two formulas we can obtain the
representation content of $\lae_{11}$ with respect to $\lae_{10}$ at
low height. 

Write a root $\alpha$ of $\lae_{11}$ as
\be
\alpha=l\alpha_{-2}+\sum_{i=-1}^8n_i\alpha_i
\ee
so that $l$ is the (hyperbolic) level of $\alpha$ with respect to $\lae_{10}$. We
can then obtain the positive part of $\lae_{11}$ as
\be
\lap(\lae_{10})_+\oplus\bigoplus_{l\ge
1}\bigoplus_{i=1}^{N_l}(L(\Lambda^{(l)}_i))^{\oplus\mu(\Lambda^{(l)}_i)},
\ee
where we assume all the $\Lambda^{(l)}_i$ occuring for a fixed $l$ to
be distinct and $\mu(\Lambda^{(l)}_i)$ is the outer multiplicity with
which they occur.

Computations using the modified Peterson and Freudenthal formulas can
be done efficiently on a computer and we present the results for the
decomposition above up to height $120$ in $\lae_{11}$ where the
maximal occuring level is $5$ in table \ref{te11e10}. 
The
computations were carried out with the help of a C program
by compiling a list of all roots of
$\lae_{11}$ to that height using the (orbit improved) 
Peterson formula and then the
Freudenthal formula to iteratively determine the relevant $\lae_{10}$
representations. 
In an appendix we list the
multiplicities of all elements in the fundamental chamber $\funddom$
for $\lae_{10}$ (up to height 340) and for $\lae_{11}$ (up to height
240) for reference.  The results we obtain agree with other data in
the literature \cite{BaeGeNi98,NiFi03}.

We already encounter representations
with outer multiplicity greater than one in this regime. At the first
level we find only the $\lae_{10}$ representation 
$V=[1,0,0,0,0,0,0,0,0,0]$ as expected from proposition \ref{ff}. This
is a highly non-trivial check on the $\lae_{10}$ and $\lae_{11}$ data and
the algorithm. The representations on the second level should be
$[V,V]/U$ where $U$ has highest weight $[0,1,0,0,0,0,0,0,0,0]$ (corresponding to the 
$\lae_{11}$ element $(2,1,0,0,0,0,0,0,0,0,0)$ which is the first 
non-trivial commutator in $[V,V]$). So we can learn someting
about the Clebsch-Gordon series of the antisymmetric product 
of the $\lae_{10}$ representation $V$ with itself at low levels 
(in the $\lae_{10}$ weight lattice). Considerations similar in spirit
apply to higher levels. Apparently, the information about tensor 
products of $\lae_{10}$ representations could also have been
obtained by other methods (character theory, say) and then been
used to generate data for $\lae_{11}$. As this does not give any
obvious technical advantage we have used the method presented here
as it makes direct use of the (possibly) more fundamental 
structure of $\lae_{11}$.\\

\end{section}

{\bf Acknowledgements:} The author would like to thank P. Goddard for
valuable 
discussions and H. Nicolai and P. West for 
comments. Support by the Studienstiftung des deutschen
Volkes, EPSRC and the Cambridge European Trust is gratefully
acknowledged.

\newpage

\begin{longtable}{ccccr}
$l$&$\lae_{11}$ height&$\lae_{10}$ highest weight
$\Lambda_i^{(l)}$&$\lae_{11}$ element&$\mu(\Lambda_i^{(l)})$\\
\hline
\hline
\endhead
1&1&[1,0,0,0,0,0,0,0,0,0]&(1,0,0,0,0,0,0,0,0,0,0)&1\\
\hline
2&35&[0,0,1,0,0,0,0,0,0,0]&(2,2,2,2,3,4,5,6,4,2,3)&1\\
&57&[0,0,0,0,0,0,0,0,0,1]&(2,3,4,5,6,7,8,9,6,3,4)&1\\
&65&[1,0,1,0,0,0,0,0,0,0]&(2,2,3,4,6,8,10,12,8,4,6)&1\\
&79&[0,1,0,0,0,0,0,0,1,0]&(2,3,4,6,8,10,12,14,9,4,7)&2\\
&87&[1,0,0,0,0,0,0,0,0,1]&(2,3,5,7,9,11,13,15,10,5,7)&1\\
&95&[2,0,1,0,0,0,0,0,0,0]&(2,2,4,6,9,12,15,18,12,6,9)&2\\
&95&[0,0,0,0,0,0,0,1,0,0]&(2,4,6,8,10,12,14,16,10,5,8)&2\\
&96&[0,1,1,0,0,0,0,0,0,0]&(2,3,4,6,9,12,15,18,12,6,9)&1\\
&98&[1,0,0,1,0,0,0,0,0,0]&(2,3,5,7,9,12,15,18,12,6,9)&2\\
&109&[1,1,0,0,0,0,0,0,1,0]&(2,3,5,8,11,14,17,20,13,6,10)&4\\
&111&[0,0,1,0,0,0,0,0,1,0]&(2,4,6,8,11,14,17,20,13,6,10)&5\\
&117&[2,0,0,0,0,0,0,0,0,1]&(2,3,6,9,12,15,18,21,14,7,10)&3\\
&118&[0,1,0,0,0,0,0,0,0,1]&(2,4,6,9,12,15,18,21,14,7,10)&4\\
\hline
3&58&[1,0,0,0,0,0,0,0,0,1]&(3,3,4,5,6,7,8,9,6,3,4)&1\\
&67&[0,1,1,0,0,0,0,0,0,0]&(3,3,3,4,6,8,10,12,8,4,6)&1\\
&73&[0,0,0,0,1,0,0,0,0,0]&(3,4,5,6,7,8,10,12,8,4,6)&1\\
&80&[1,1,0,0,0,0,0,0,1,0]&(3,3,4,6,8,10,12,14,9,4,7)&1\\
&82&[0,0,1,0,0,0,0,0,1,0]&(3,4,5,6,8,10,12,14,9,4,7)&2\\
&88&[2,0,0,0,0,0,0,0,0,1]&(3,3,5,7,9,11,13,15,10,5,7)&1\\
&89&[0,1,0,0,0,0,0,0,0,1]&(3,4,5,7,9,11,13,15,10,5,7)&3\\
&95&[1,0,0,0,0,0,0,0,2,0]&(3,4,6,8,10,12,14,16,10,4,8)&2\\
&96&[1,0,0,0,0,0,0,1,0,0]&(3,4,6,8,10,12,14,16,10,5,8)&3\\
&97&[1,1,1,0,0,0,0,0,0,0]&(3,3,4,6,9,12,15,18,12,6,9)&2\\
&99&[2,0,0,1,0,0,0,0,0,0]&(3,3,5,7,9,12,15,18,12,6,9)&2\\
&99&[0,0,2,0,0,0,0,0,0,0]&(3,4,5,6,9,12,15,18,12,6,9)&1\\
&100&[0,1,0,1,0,0,0,0,0,0]&(3,4,5,7,9,12,15,18,12,6,9)&4\\
&103&[1,0,0,0,1,0,0,0,0,0]&(3,4,6,8,10,12,15,18,12,6,9)&5\\
&104&[0,0,0,0,0,0,0,0,1,1]&(3,5,7,9,11,13,15,17,11,5,8)&4\\
&108&[0,0,0,0,0,1,0,0,0,0]&(3,5,7,9,11,13,15,18,12,6,9)&3\\
&110&[2,1,0,0,0,0,0,0,1,0]&(3,3,5,8,11,14,17,20,13,6,10)&3\\
&111&[0,2,0,0,0,0,0,0,1,0]&(3,4,5,8,11,14,17,20,13,6,10)&6\\
&112&[1,0,1,0,0,0,0,0,1,0]&(3,4,6,8,11,14,17,20,13,6,10)&12\\
&115&[0,0,0,1,0,0,0,0,1,0]&(3,5,7,9,11,14,17,20,13,6,10)&11\\
&118&[3,0,0,0,0,0,0,0,0,1]&(3,3,6,9,12,15,18,21,14,7,10)&3\\
&119&[1,1,0,0,0,0,0,0,0,1]&(3,4,6,9,12,15,18,21,14,7,10)&14\\
\hline
4&79&[0,0,0,0,0,1,0,0,0,0]&(4,5,6,7,8,9,10,12,8,4,6)&1\\
&83&[1,0,1,0,0,0,0,0,1,0]&(4,4,5,6,8,10,12,14,9,4,7)&1\\
&90&[1,1,0,0,0,0,0,0,0,1]&(4,4,5,7,9,11,13,15,10,5,7)&1\\
&92&[0,0,1,0,0,0,0,0,0,1]&(4,5,6,7,9,11,13,15,10,5,7)&2\\
&97&[2,0,0,0,0,0,0,1,0,0]&(4,4,6,8,10,12,14,16,10,5,8)&1\\
&97&[0,1,0,0,0,0,0,0,2,0]&(4,5,6,8,10,12,14,16,10,4,8)&1\\
&98&[0,1,0,0,0,0,0,1,0,0]&(4,5,6,8,10,12,14,16,10,5,8)&2\\
&99&[0,2,1,0,0,0,0,0,0,0]&(4,4,4,6,9,12,15,18,12,6,9)&1\\
&101&[1,1,0,1,0,0,0,0,0,0]&(4,4,5,7,9,12,15,18,12,6,9)&2\\
&103&[0,0,1,1,0,0,0,0,0,0]&(4,5,6,7,9,12,15,18,12,6,9)&2\\
&104&[2,0,0,0,1,0,0,0,0,0]&(4,4,6,8,10,12,15,18,12,6,9)&1\\
&105&[0,1,0,0,1,0,0,0,0,0]&(4,5,6,8,10,12,15,18,12,6,9)&5\\
&105&[1,0,0,0,0,0,0,0,1,1]&(4,5,7,9,11,13,15,17,11,5,8)&5\\
&109&[1,0,0,0,0,1,0,0,0,0]&(4,5,7,9,11,13,15,18,12,6,9)&5\\
&112&[1,2,0,0,0,0,0,0,1,0]&(4,4,5,8,11,14,17,20,13,6,10)&2\\
&113&[2,0,1,0,0,0,0,0,1,0]&(4,4,6,8,11,14,17,20,13,6,10)&5\\
&113&[0,0,0,0,0,0,0,1,1,0]&(4,6,8,10,12,14,16,18,11,5,9)&3\\
&114&[0,1,1,0,0,0,0,0,1,0]&(4,5,6,8,11,14,17,20,13,6,10)&9\\
&114&[0,0,0,0,0,0,0,0,0,2]&(4,6,8,10,12,14,16,18,12,6,8)&2\\
&115&[0,0,0,0,0,0,1,0,0,0]&(4,6,8,10,12,14,16,18,12,6,9)&7\\
&116&[1,0,0,1,0,0,0,0,1,0]&(4,5,7,9,11,14,17,20,13,6,10)&14\\
&120&[2,1,0,0,0,0,0,0,0,1]&(4,4,6,9,12,15,18,21,14,7,10)&5\\
&120&[0,0,0,0,1,0,0,0,1,0]&(4,6,8,10,12,14,17,20,13,6,10)&10\\
\hline
5&101&[0,0,1,0,0,0,0,1,0,0]&(5,6,7,8,10,12,14,16,10,5,8)&1\\
&106&[1,1,0,0,1,0,0,0,0,0]&(5,5,6,8,10,12,15,18,12,6,9)&1\\
&106&[2,0,0,0,0,0,0,0,1,1]&(5,5,7,9,11,13,15,17,11,5,8)&1\\
&107&[0,1,0,0,0,0,0,0,1,1]&(5,6,7,9,11,13,15,17,11,5,8)&1\\
&108&[0,0,1,0,1,0,0,0,0,0]&(5,6,7,8,10,12,15,18,12,6,9)&1\\
&111&[0,1,0,0,0,1,0,0,0,0]&(5,6,7,9,11,13,15,18,12,6,9)&2\\
&114&[1,0,0,0,0,0,0,1,1,0]&(5,6,8,10,12,14,16,18,11,5,9)&1\\
&115&[1,1,1,0,0,0,0,0,1,0]&(5,5,6,8,11,14,17,20,13,6,10)&2\\
&115&[1,0,0,0,0,0,0,0,0,2]&(5,6,8,10,12,14,16,18,12,6,8)&2\\
&116&[1,0,0,0,0,0,1,0,0,0]&(5,6,8,10,12,14,16,18,12,6,9)&3\\
&117&[2,0,0,1,0,0,0,0,1,0]&(5,5,7,9,11,14,17,20,13,6,10)&2\\
&117&[0,0,2,0,0,0,0,0,1,0]&(5,6,7,8,11,14,17,20,13,6,10)&2\\
&118&[0,1,0,1,0,0,0,0,1,0]&(5,6,7,9,11,14,17,20,13,6,10)&4\\
\caption{\label{te11e10} $\lae_{10}$ representations occuring in
$\lae_{11}$ up to height $120$.}
\end{longtable}

\newpage

\appendix

\begin{section}{$\lae_{10}$ up to height $340$}

The following table contains the multiplicities of all elements
$\beta$ in the
fundamental chamber of $\lae_{10}$ up to height $340$. As noted
before \cite{GeNi95}, the
discrepancy between the multiplicity of a root $\beta$ and the 
number of transverse polarisations $p_8(1-\half\beta^2)$, 
generated by the partition function on 8 colours 
$\sum_np_8(n)q^n=\prod_{n\ge 1} (1-q^n)^{-8}$, grows with increasing
$-\beta^2$  and 
$\mult(\beta)\ge p_8(1-\half\beta^2)$. The results of
\cite{GeNi95,BaeGe97} show that the discrepancy is not only due to
additional longitudinal states but that also some transversal states
disappear from the spectrum. Furthermore, the anisotropy grows
with increasing height, i.e. the amount by which the multiplicity fails
to be a function just of $\beta^2$. 

{\small
\begin{longtable}{ccccr}
$ht(\beta)$&Fund. weight basis&Simple root basis&$\beta^2$&$\mult(\beta)$\\
\hline
\endhead
30&[1,0,0,0,0,0,0,0,0,0]&(0,1,2,3,4,5,6,4,2,3)&0&8\\
60&[2,0,0,0,0,0,0,0,0,0]&(0,2,4,6,8,10,12,8,4,6)&0&8\\
61&[0,1,0,0,0,0,0,0,0,0]&(1,2,4,6,8,10,12,8,4,6)&-2&44\\
76&[0,0,0,0,0,0,0,0,1,0]&(2,4,6,8,10,12,14,9,4,7)&-4&192\\
90&[3,0,0,0,0,0,0,0,0,0]&(0,3,6,9,12,15,18,12,6,9)&0&8\\
91&[1,1,0,0,0,0,0,0,0,0]&(1,3,6,9,12,15,18,12,6,9)&-4&192\\
93&[0,0,1,0,0,0,0,0,0,0]&(2,4,6,9,12,15,18,12,6,9)&-6&727\\
106&[1,0,0,0,0,0,0,0,1,0]&(2,5,8,11,14,17,20,13,6,10)&-8&2472\\
\end{longtable}}

\end{section}


\begin{section}{$\lae_{11}$ up to height $240$}

This table contains the multiplicities of all elements $\beta$ in the
fundamental domain of $\lae_{11}$ up to height $240$. Unlike in the hyperbolic
$\lae_{10}$ case not all $\beta$ have connected support and so are
imaginary roots of the algebra. We list only the non-trivial elements.
Again there is anisotropy but in stark contrast to the $\lae_{10}$
case now all multiplicities are (almost always strictly) 
less than the number of transverse
polarisations: $\mult(\beta)\le p_9(1-\half\beta^2)$.

{\small
\begin{longtable}{ccccr}
$ht(\beta)$&Fund. weight basis&Simple root basis&$\beta^2$&$\mult(\beta)$\\
\hline
\endhead
30&[0,1,0,0,0,0,0,0,0,0,0]&(0,0,1,2,3,4,5,6,4,2,3)&0&8\\
47&[0,0,0,0,0,0,0,0,0,1,0]&(1,2,3,4,5,6,7,8,5,2,4)&-2&46\\
60&[0,2,0,0,0,0,0,0,0,0,0]&(0,0,2,4,6,8,10,12,8,4,6)&0&8\\
61&[1,0,1,0,0,0,0,0,0,0,0]&(0,1,2,4,6,8,10,12,8,4,6)&-2&44\\
64&[0,0,0,1,0,0,0,0,0,0,0]&(1,2,3,4,6,8,10,12,8,4,6)&-4&206\\
76&[2,0,0,0,0,0,0,0,0,1,0]&(0,2,4,6,8,10,12,14,9,4,7)&-4&192\\
77&[0,1,0,0,0,0,0,0,0,1,0]&(1,2,4,6,8,10,12,14,9,4,7)&-6&801\\
86&[1,0,0,0,0,0,0,0,0,0,1]&(1,3,5,7,9,11,13,15,10,5,7)&-8&2801\\
\end{longtable}}
\end{section}

\end{document}